\documentclass{article}
\usepackage{spconf}
\usepackage[utf8]{inputenc}   % UNIX, codage ISO 8859-1
\usepackage[english]{babel}   % "babel.sty" + "french.sty"
\usepackage{times}			% ajout times le 30 mai 2003
\usepackage{bm}
\usepackage{enumitem}
\usepackage{amsmath}
\usepackage{amssymb}
\usepackage{amsthm}
\usepackage{amsfonts}
\usepackage{hyperref}
\usepackage{algorithm, algorithmic}
\usepackage{graphicx}

  \newcommand{\bL}{\textbf{L}}
  \newcommand{\bG}{\textbf{G}}
  \newcommand{\bA}{\textbf{A}}
  \newcommand{\bD}{\textbf{D}}

  \newcommand{\bM}{\textbf{M}}
  \newcommand{\bI}{\textbf{I}}

  \newcommand{\br}{\textbf{r}}

  \newcommand{\G}{\mathcal{G}}
  
    \newcommand{\V}{\mathcal{V}}
    \newcommand{\E}{\mathbb{E}}
    \newcommand{\T}{\mathcal{T}}

\renewcommand{\O}{\mathcal{O}}
    
\DeclareMathOperator*{\Tr}{Tr}
\DeclareMathOperator*{\argmin}{\mathrm{argmin}}

\title{Estimating the inverse trace using random forests on graphs}

\name{S. Barthelmé (1), N. Tremblay (1), A. Gaudillière (2), L.
  Avena (3) and
  P-O Amblard (1)}
\address{(1) Univ Grenoble-Alpes, CNRS, Grenoble-INP, GIPSA-lab, Grenoble,
  France \\
  (2) Aix-Marseille Univ, CNRS, I2M, Marseille, France \\
  (3) Leiden University, Netherlands
}

\begin{document}
       \maketitle
       \begin{abstract}
         Some data analysis problems require the computation of (regularised) inverse traces, i.e. quantities of the form  $\Tr (q \bI + \bL)^{-1}$. For large matrices, direct methods are unfeasible and one must resort to approximations, for example using a conjugate gradient solver combined with Girard's trace estimator (also known as Hutchinson's trace estimator). Here we describe an unbiased estimator of the regularized inverse trace, based on Wilson's algorithm, an algorithm that was initially designed to draw uniform spanning trees in graphs. Our method is fast,  easy to implement, and scales to very large matrices. Its main drawback is that it is limited to diagonally dominant matrices $\bL$.
     \end{abstract}
\vspace{.5cm}
     
%\section{Background}

Monte Carlo methods are increasingly popular in large-scale linear algebra problems \cite{mahoney2011randomized}. Among the many different quantities one may need to compute on large matrices, spectral summaries of the form 
$ \sum_{i=1}^{n} f(\lambda_{i}(\bL)) $,
where the $\lambda_{i}$'s are the eigenvalues of $\bL$ and $f$ is some function, are often required. Here we focus on the following quantity:
\begin{equation}
  \label{eq:invtrace}
  s(q) = q \Tr((\bL + q\bI)^{-1}) = \sum_{i=1}^n \frac{q}{\lambda_i+q}
\end{equation}
which we seek to evaluate for real $q>0$. We call the quantity $s(q)$ because it is equivalent (up to scaling) to the Stieltjes transform of the eigenvalue density evaluated on the negative real axis \cite{anderson2010introduction}.
%We assume that $\bL$ is positive definite, in which case $s(q)$ can be seen to measure the degrees of freedom of linear smoothing by $\bL$.

In practice, the problem of estimating efficiently $s(q)$ may arise when looking for the optimal regularization parameter in a regularized optimization problem. Say we measure a signal $\mathbf{x} = [x_{1}, \ldots, x_{n}]^{t}$ under white Gaussian noise $\epsilon$. The measurements read $y_{i} = x_i + \epsilon_i$ for $i=1$ to $n$. Many estimation methods (smoothing splines, semi-supervised learning, Gaussian process regression) define an estimator of $\mathbf{x}$ as:
\begin{equation}
\label{eq:linear_estimation}
\hat{\mathbf{x}} = \argmin_{\mathbf{z}\in\mathbb{R}^n} \frac{q}{2} \left|\left| \mathbf{y} - \mathbf{z} \right| \right|^{2} + \frac{1}{2} \mathbf{z}^{t} \bL \mathbf{z}
\end{equation}
where $\bL$ is a semi-definite positive matrix defining the penalty (regularisation) term, and $q$ parametrizes the regularisation's strength. The solution to this optimisation problem equals:
\begin{equation}
\label{eq:linear_estimation_sol}
\hat{\mathbf{x}} = q (q \bI + \bL)^{-1} \mathbf{y}.
\end{equation}
In most cases the optimal value of $q$ is unknown and must be estimated, for instance using AIC (Akaike's Information Criterion) or Generalised-Cross Validation (GCV). AIC requires computing the number of degrees of freedom of the estimator, which here can be taken to equal $s(q)$ (see \cite{ESL}, ch. 5, \cite{girard1987algorithme,girard1989fast}). 

The simplest solution to compute eq. (\ref{eq:invtrace}) is of course to compute the eigenvalues of $\bL$, which comes at $\O(n^3)$ cost if $\bL$ is $n \times n$. Moreover, there is no particular gain to expect from the sparsity of $\bL$. In fact, iterative methods for eigenvalues, that look to estimate the smallest or largest eigenvalues of $\bL$, cannot be used directly here, as $s(q)$ involves the whole spectral density. 
An alternative is to consider Monte Carlo methods. A famous estimator for the trace of a matrix was first suggested by Girard in~\cite{girard1987algorithme}: let $\br$ denote a length-$n$ vector of independent, standard Gaussian entries. 
Let $\bM$ denote a $n \times n$ matrix. Then:
\begin{equation}
  \label{eq:girard}
  \E( \br^{t} \bM \br) =  \E \left(  \Tr (\bM \br \br^{t}) \right) =
  \Tr  \left(\bM \E (\br \br^{t})\right) = \Tr \bM
\end{equation}
%This leads immediately to the idea of estimating $\Tr \bM$ via an empirical average:
This leads immediately to estimating $\Tr \bM$ using the empirical mean $\Tr \bM \approx \frac{1}{k} \sum_{l=1}^{k} \br_{l}^{t} \bM \br_{l}$.
%\begin{equation}
 % \label{eq:girard-empirical}
 % \Tr \bM \approx \frac{1}{k} \sum_{l=1}^{k} \br_{l}^{t} \bM \br_{l}.
%\end{equation}
Note that eq. (\ref{eq:girard}) is valid for any random vector  with diagonal covariance, so we may use other random vectors \cite{hutchinson1990stochastic}. Various options have been studied in the literature, see \cite{avron_randomized_2011-1}. In this work we use Gaussian vectors for simplicity (as we will see, it is not the main factor here). 

%In our case, $\bM = q( q\bI + \bL)^{-1}$, and the unbiased Girard estimator of $s(q)$ using $k$ random vectors reads
In our case, $\bM = q( q\bI + \bL)^{-1}$, and   Girard's estimator of $s(q)$ reads
\begin{align}
\label{eq:girard-estimator}
\hat{s}^\texttt{G}_k(q)= \frac{q}{k}\sum_{l=1}^k \br_{l}^{t}(q\bI + \bL)^{-1}\br_{l}.
\end{align} 
In the Gaussian case, the variance of the estimation for $k=1$~(see, e.g.,  lemma 9 of~\cite{avron_randomized_2011-1}) is:
\begin{align}
\label{eq:girard-gaussian-variance}
\text{Var}(\hat{s}^\texttt{G}_1(q)) = \sum_{i=1}^n \frac{2q^2}{(q+\lambda_i)^2}.
\end{align}
We still need to figure out how to compute the quadratic forms $\br^{t} (q\bI + \bL)^{-1} \br$ in eq.~\eqref{eq:girard-estimator}. This involves solving a large linear system, a task for which algorithms abound. If $\bL$ is sparse, computing a sparse Cholesky factor will give good results for many systems, up to a certain size \footnote{In fact, if the Cholesky factor is available, the Takahashi equations may also be used to obtain the trace, see \cite{rue2005gaussian}}. Alternatively, for very large systems, iterative solvers such as Conjugate Gradients may be used \cite{barrett1994templates}. Another approach is to use an order $p$ polynomial approximation\footnote{Using Chebychev polynomials for instance if one wants to ensure the smallest infinite-norm error:  $\text{sup}_{x\in[0, \lambda_{\text{max}}]}\;|f(x)-\sum_{j=0}^p \alpha_j x^j|$}  of the function $f(x)=q/(q+x)\simeq\sum_{j=0}^p \alpha_j x^j$. Estimating  $\br^{t} (q\bI + \bL)^{-1} \br$ then boils down to computing $\br^{t} \sum_{j=0}^p \alpha_j \bL^j \br$, that is: $p$ matrix vector multiplications and one scalar product. Iterative solvers and polynomial methods only provide  approximate solutions, but we expect the error induced by these approximations to be small relative to the Girard variance of Eq.~\eqref{eq:girard-gaussian-variance}. A combination of Girard's trace estimator and iterative solvers has been used, e.g., in \cite{stein2013stochastic}.

Below, we describe an alternative method that is very natural and intrinsic when $\bL$ is actually a graph Laplacian, a particular class of matrices associated with graphs. At the end of section 2 we extend the technique to diagonally dominant matrices, {\it i.e.} the set of matrices that verify $\forall i\quad L_{ii} \geq \sum_{j\neq i} \left| L_{ij} \right|$.
%Section \ref{sec:SDD-matrices} extends our technique to diagonally dominant matrices, i.e. the set of matrices that verify:
%\begin{equation}
%  \label{eq:diag-dom}
%  \forall i\quad L_{ii} \geq \sum_{j\neq i} \left| L_{ij} \right|.
%\end{equation}
\vspace{.3cm}

\noindent {\large \bf 2 \quad Uniform spanning trees, random forests, and inverse traces}
\vspace{.2cm}

%\section{Uniform spanning trees, random forests, and inverse traces}
%\label{sec:USTs}

In this section we recall some facts on graphs and spanning trees that should help understand our method. %We state all results on unweighted graphs for simplicity but all results carry over to the weighted case.
Mathematical details can be found in \cite{avena2013some} and \cite{avena_two_2017}. 

Consider a weighted graph $\mathcal{G}=(\V,\mathcal{E})$ of $n=|\V|$ nodes and $|\mathcal{E}|$ edges. We restrict ourselves to undirected graphs in this paper, even though the results may be extended to strongly connected\footnote{Given any pair of nodes $(i,j)$, there is a directed path to go from $i$ to $j$, \textit{and} from $j$ to $i$.} directed graphs. We denote by $\bA\in\mathbb{R}^{n\times n}$ the graph's adjacency matrix, where $A_{ij}=A_{ji}\geq0$ is the weight of the connection between nodes $i$ and $j$. The graph Laplacian of $\G$ equals $\bL = \bD - \bA\in\mathbb{R}^{n\times n}$,
%\begin{equation}
%  \label{eq:graph_laplacian}
%  \bL = \bD - \bA\in\mathbb{R}^{n\times n},
%\end{equation} 
where $\bD=\text{diag}(d_1, \ldots, d_n)\in\mathbb{R}^{n\times n}$ is the diagonal degree matrix with $d_i=\sum_{j} A_{ij}$ the degree of node $i$.
The graph Laplacian is a fascinating object with many applications in machine learning and graph signal processing, see eg.~\cite{chung_complex_2006}.

%Let $\G = (\V,\mathcal{E})$ denote an undirected graph with $n = |\V|$ vertices and $|\mathcal{E}|$ edges. We denote by $\bA\in\mathbb{R}^{n\times n}$ its adjacency matrix: $A_{ij}=A_{ji}$ is equal to zero if there is no edge betwwen nodes $i$ and $j$, and is equal to the weight of the link connecting node $i$ to $j$, and by $\bD\in\mathbb{R}^{n\times n}$ its degree matrix: a diagonal matrix where $D_{ii}=d_i$ equals the number of neighbours of vertex $i$. The graph Laplacian of $\G$ equals:
%\begin{equation}
%  \label{eq:graph_laplacian}
%  \bL = \bD - \bA\in\mathbb{R}^{n\times n}.
%\end{equation}

A tree is a cycle-free graph, and a spanning tree $\T$ of $\G$ is a cycle-free connected subgraph of $\G$ that spans all $n$ nodes of $\G$. A typical graph has more than one spanning tree. For instance, the complete graph of size $n$ contains $n^{n-2}$ different spanning trees. A tree sampled uniformly from the set of all spanning trees of $\G$ is called a uniform spanning tree (UST).

A fast algorithm for sampling USTs, now known as "Wilson's algorithm" was developed in \cite{wilson1996generating} . In a nutshell, the algorithm runs as follows: pick a node at random, and call it the root of the tree. Now pick another node, and run a random walk until it hits the root. The trajectory of the random walk may include loops: we simply erase them as they come. The resulting ``loop-erased'' random walk will form the first branch of the spanning tree. Next, pick a node that is not yet in the tree, run a random walk until it hits the tree, erase the possible loops, add this new branch to the tree, etc. Wilson's algorithm runs in time proportional to $\O(\tau )$ where $\tau$ is the average ``commute time'': the time it takes a random walk to reach node $j$ starting from node $i$ for two nodes picked uniformly on the graph.

\begin{algorithm}[tb]
	\caption{A variant of Wilson's algorithm}
	\label{alg:modified_wilson}
	\begin{algorithmic}
	\STATE \textbf{Input:} A graph $\mathcal{G}=(\V, \mathcal{E})$ of size $n$ and $q>0$\\
	$\mathcal{R}\leftarrow \emptyset$, $\mathcal{W}\leftarrow \emptyset$\\
	Add a node, called $\Delta$, to $\mathcal{G}$ and connect it to all $n$ nodes in $\V$ with edges of weight $q$. Call this augmented graph $\mathcal{G}'$.\\
	\textbf{while} $\mathcal{W}\neq\V$ \textbf{do}:\\
	$\quad\bm{\cdot}$ Do a random walk on $\mathcal{G}'$ starting from any node $i\in\V\setminus\mathcal{W}$ until it reaches either $\Delta$, or a node in $\mathcal{W}$.\\
	$\quad\bm{\cdot}$ Erase all the loops of the trajectory, in the order of appearance.\\
	$\quad\bm{\cdot}$ Add all the nodes of this loop-erased trajectory to the set of visited nodes $\mathcal{W}$. \\
	$\quad$\textbf{if} the last node of the trajectory is $\Delta$ \textbf{do}: \\
	$\qquad\bm{\cdot}$ Denote by $l$ the last visited node before $\Delta$\\
	$\qquad\bm{\cdot}$ $\mathcal{R}\leftarrow \mathcal{R}\cup\{l\}$ \\
	\textbf{Output:} $\mathcal{R}$, the root set of the sampled  forest spanning $\mathcal{G}$.
	\end{algorithmic}
\end{algorithm}

Wilson in \cite{wilson1996generating} noted that his algorithm could be used to generate random spanning forests, and not just USTs. A forest is a set of trees, and a spanning forest is a set of disjoint trees that, taken together, span the whole graph. The algorithm\footnote{Alg.~\ref{alg:modified_wilson} is written in order to only output the set of roots of the sampled forest, as this is the information we will use in this paper. Much more information can in practice be extracted.} is given as alg.~\ref{alg:modified_wilson}: it uses loop-erased random walks (LERW), but these LERWs may be interrupted early. At each node, the random walk is interrupted with probability $\frac{q}{q+d_i}$. Of course, the larger $q$, the shorter the walks, the larger the number of roots, the faster the algorithm. In the implementation given in alg.~\ref{alg:modified_wilson}, the average runtime is\footnote{This figure assumes that, when at node $i$, picking a neighbour at random is $\O(d_i)$. This can be marginally improved by some preprocessing tricks, for example by using the alias method for sampling. In addition, in the case of unweighted graphs there is no dependency on the degree (picking a random neighbor is $\O(1)$) } $\O(|E|/q)$. 

The resulting process has many fascinating aspects, some of which have been investigated in~\cite{avena2018random}. For our purposes, we focus on the fact that the number of roots is in fact an unbiased estimator of $s(q)$: 
\begin{align}
\E(|\mathcal{R}|) = \sum_{i=1}^n \frac{q}{q+\lambda_i}=s(q).
\end{align}
%Running alg.~\ref{alg:modified_wilson} $k$ times, one obtains $k$ sets of roots $\{\mathcal{R}_l\}_{l=1,\ldots,k}$. 
This suggests to define  Wilson estimator of $s$ as:
\begin{equation}
  \label{eq:number-of-roots}
  \hat{s}^\texttt{W}_k(q) = \frac{1}{k}\sum_{l=1}^k |\mathcal{R}_l|.
\end{equation}
where the $k$ sets of roots $\{\mathcal{R}_l\}_{l=1,\ldots,k}$ are obtained by running alg.~\ref{alg:modified_wilson} $k$ times.
%Our suggestion is therefore quite simple: in order to estimate $s(q)$, run alg.~\ref{alg:modified_wilson} $k$ times, and compute the average number of roots. 
A further property of alg.~\ref{alg:modified_wilson} is, in the case $k=1$ (see~\cite{avena2018random}):
\begin{equation}
  \label{eq:number-of-roots}
  \text{Var}(\hat{s}^\texttt{W}_1(q)) = q \sum_{i=1}^{n} \frac{\lambda_i}{\left( q+ \lambda_{i} \right)^{2}}.
\end{equation}
This variance can be compared with Girard's (eq.~\eqref{eq:girard-gaussian-variance}): we see that for both very small and very large values of $q$, Girard's estimator is less effective per sample. Unfortunately, identifying exactly the interval of $q$ for which Wilson's estimator is preferable (on a per-sample basis) is heavily dependent on the eigenvalue distribution.

%One can check easily that $\text{Var}(\hat{s}^\texttt{W}_1) \leq \E(\hat{s}^\texttt{W}_1)$, so that the relative error after $k$ realisations verifies (as $s(q)\geq 1$):
Since $\text{Var}(\hat{s}^\texttt{W}_1) \leq \E(\hat{s}^\texttt{W}_1)$ and $s(q)\geq 1$ the relative error verifies: 
\begin{equation}
  \label{eq:rel-err}
   \frac{\text{Var}(\hat{s}^\texttt{W}_k)}{\E(\hat{s}^\texttt{W}_k)^{2}}=
  \frac{\text{Var}(\hat{s}^\texttt{W}_1)}{k\E(\hat{s}^\texttt{W}_1)^{2}}  \leq    \frac{1}{k s(q) } \leq \frac{1}{k}.
\end{equation}
Let us point out several advantages of the suggested algorithm. First, no preprocessing is required. The graph does even not need to be pre-computed: essentially, all we need is the ability to run a random walk on the graph. Second, it is very easy to implement (our implementation runs under 20 lines of Julia code). Third, it is is easy to parallelise, as we can just generate several forests concurrently. Fourth, its memory footprint is minimal, requiring a handful of $\O(n)$ quantities. 
However, the main disadvantage is that the algorithm can only estimate $s(q)$ if $\bL$ is a graph Laplacian. The next section partly lifts that restriction to allow the use of diagonally-dominant matrices.
\vspace{.1cm}

\noindent {\bf Generalising to diagonally-dominant matrices.}
%\label{sec:SDD-matrices}
We borrow a trick from the rich literature on Laplacian solvers (see for instance~\cite{kelner2013simple,hunter2014computing}). Let $\bG$ be a diagonally dominant matrix, that we decompose as $\bG = \mathbf{D}_{1} + \mathbf{D}_{2} + \bA_{p} + \bA_{n}$
%\begin{equation}
 % \label{eq:decomposition}
 % \bG = \mathbf{D}_{1} + \mathbf{D}_{2} + \bA_{p} + \bA_{n},
%\end{equation}
where:
\begin{itemize}
\item $\bA_{p}$ contains the positive off-diagonal elements, $\bA_{n}$ contains the negative ones
\item $\mathbf{D}_{1}$ is a diagonal matrix, with $D_{1}(i,i) = \sum_{j \neq i} |G_{ij}|$ (sum of off-diagonal elements)
\item $\mathbf{D}_{2}$ is also diagonal, with entries $D_{2}(i,i) = G_{ii}-D_{1}(i,i)$. Diagonal dominance of $\bG$ implies that $\forall i, D_2(i,i)\geq 0$. 
\end{itemize}
In the same way we restricted the previous discussion to undirected graphs, we here restrict ourselves to symmetric diagonally dominant matrices, implying that $\bA_{p}$ and $\bA_{n}$ are symmetric. 
%We form the following two graph Laplacians, both representing undirected weighted graphs. The first one is of size $n$:
%\begin{equation}
%  \label{eq:extended_graph_lap1}
%  \bL_{1} =    \mathbf{D}_{1} + \bA_{n} - \bA_{p}.
%\end{equation}
%while the second one is of size $2n$:
%\begin{equation}
%  \label{eq:extended_graph_lap2}
%  \bL_{2} =
%  \begin{pmatrix}
%    \mathbf{D}_{1} + \mathbf{D}_{2}/2 + \bA_{n} & -\mathbf{D}_{2}/2 - \bA_{p} \\
%    -\mathbf{D}_{2}/2 - \bA_{p} & \mathbf{D}_{1} + \mathbf{D}_{2}/2 + \bA_{n} 
%  \end{pmatrix}.
%\end{equation}
We form the following two graph Laplacians, both representing undirected weighted graphs, and of respective size $n$ and $2n$:\begin{eqnarray}
 \bL_{1} &= &  \mathbf{D}_{1} + \bA_{n} - \bA_{p} \\ 
 \bL_{2} &=&
  \begin{pmatrix}
    \mathbf{D}_{1} + \mathbf{D}_{2}/2 + \bA_{n} & -\mathbf{D}_{2}/2 - \bA_{p} \\
    -\mathbf{D}_{2}/2 - \bA_{p} & \mathbf{D}_{1} + \mathbf{D}_{2}/2 + \bA_{n} 
  \end{pmatrix}.
\end{eqnarray}
It can be easily verified that an eigenvector basis for $\bL_{2}$ can be constructed as follows: $n$ eigenvectors of the form $
\begin{pmatrix}
  \mathbf{x} \\ \mathbf{x}
\end{pmatrix}
$, where $\mathbf{x}$ is an eigenvector of $\bL_{1}$; and $n$ other eigenvectors of the form $
\begin{pmatrix}
  \mathbf{y} \\ \mathbf{-y}
\end{pmatrix}
$,  where $\mathbf{y}$ is an eigenvector of $\bG$. This implies that $\lambda(\bL_{2}) = \lambda(\bL_{1}) \cup \lambda(\bG)$
%\begin{equation}
  %\label{eq:union_eigenvalues}
  %\lambda(\bL_{2}) = \lambda(\bL_{1}) \cup \lambda(\bG)
%\end{equation}
and consequently that:
\begin{equation}
  \label{eq:diff_trace}
  s_{\bG}(q) = s_{\bL_{2}}(q) - s_{\bL_{1}}(q).
\end{equation}
Given eq. (\ref{eq:diff_trace}), the extension to symmetric diagonally dominant matrices is thus straightforward: form the two Laplacians $\bL_{1}$ and $\bL_{2}$, run the algorithm on each graph, and subtract. 
\vspace{.3cm}

\noindent {\large \bf 3 \quad Empirical results}
\vspace{.2cm}
%\section{Empirical Results}
%\label{sec:results}

We implemented our algorithm in the Julia programming language\footnote{\url{julialang.org}}, and compared its performance on a number of graphs to alternatives based on Girard's estimator. We ran all algorithms on a single core on a desktop PC. 
Specifically, the alternative algorithms are as follows. First generate $k$ Gaussian vectors of size $n$, of zero mean and variance $1$, then compute $\hat{s}_k^\texttt{G}(q)=(q/k) \sum_{l=1}^k \mathbf{r}_{l}^{t} ( \bL + q\bI)^{-1} \mathbf{r}_{l}^{t}$ using one of the following methods:
\begin{enumerate}
\item \texttt{direct}: use Julia's backslash operator (which calls CHOLMOD internally)
\item \texttt{amg}: Algebraic Multigrid (AMG) with Ruge-Stüben coarsening \cite{ruge1987algebraic}, implemented in the AlgebraicMultigrid package \footnote{https://github.com/JuliaLinearAlgebra/AlgebraicMultigrid.jl}
\item \texttt{cg}: Conjugate Gradients: we used the implementation in the IterativeSolvers.jl package \footnote{https://juliamath.github.io/IterativeSolvers.jl/dev/},  % with either diagonal preconditioning
with  diagonal preconditioning
\item \texttt{cg-amg}: same as above, with AMG preconditioning
%\item \texttt{poly}: Chebychev polynomial approximation.
\end{enumerate}

\begin{figure*}[htb]
	\begin{center}
		\includegraphics[width=0.8\textwidth]{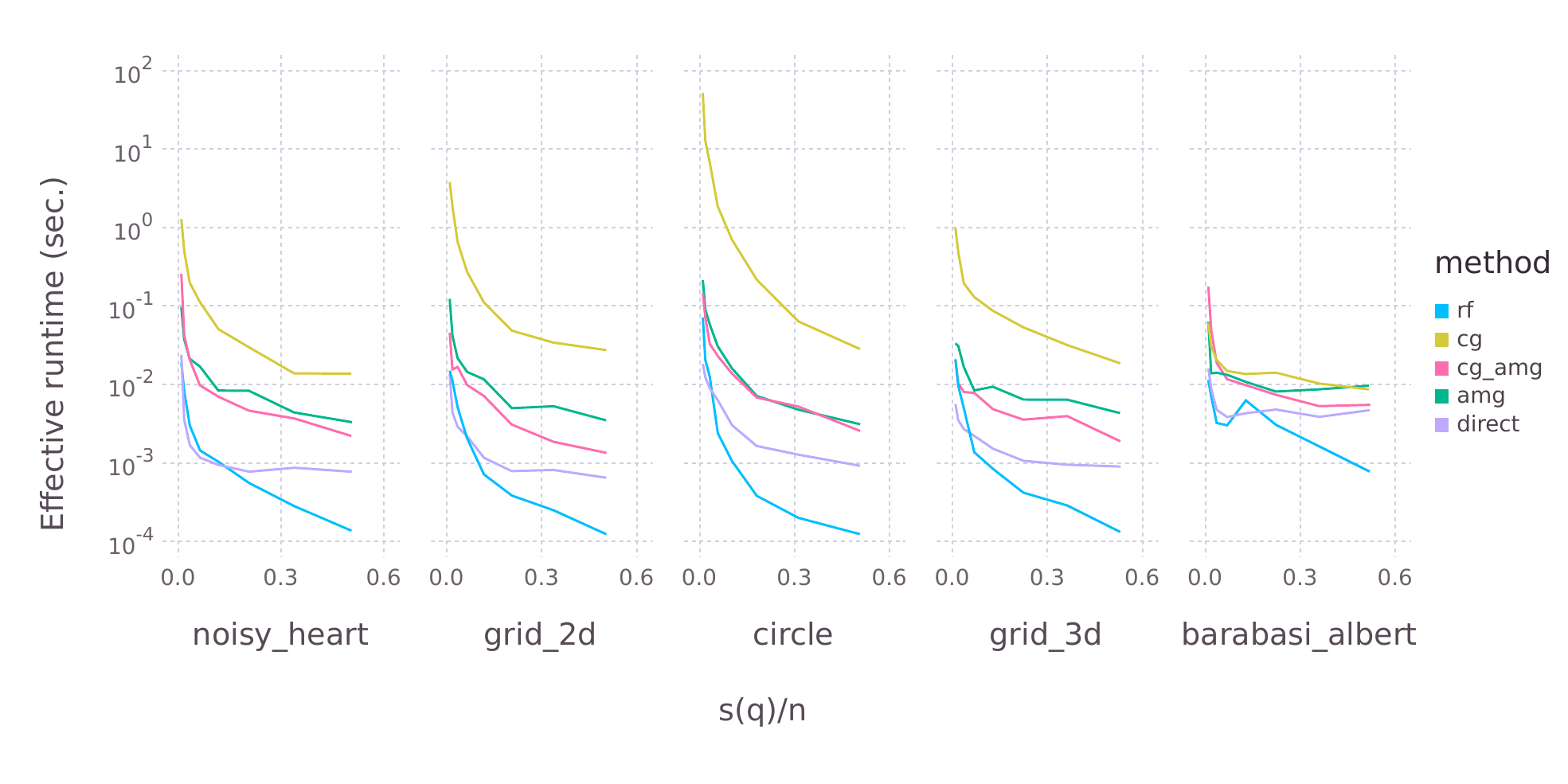}
    	\caption{ Runtime of the proposed method (``\texttt{rf}'') compared to
    alternatives, on 5 graphs. See text for details. }

	\end{center}
	\label{fig:results}
\end{figure*}

All methods defined here are based on Monte Carlo, and have an asymptotic relative error of $\epsilon^2=\text{Var}(\hat{s}_1)/k$. In order to ensure a fair comparison, we report effective runtimes as the time needed per iteration multiplied by the number of iterations needed in order to reach a fixed relative error $\epsilon$. For each value of $q$, we run each method $100$ times on each graph. This gives us an estimate $\hat{s}_{100}(q)$, along with an estimated standard deviation $\hat{\sigma}_{s(q)}$. The asymptotic relative error is given by:
\begin{equation}
\label{eq:1}
\epsilon = \frac{\hat{\sigma}_{s(q)}}{\hat{s}(q) \sqrt{k}}.
%k = \frac{r^{2}}{\sigma^{2}s^{2}}
\end{equation}
We solve for $k$ given a relative error of $\epsilon=0.02$. The time per iteration is then computed as the total time divided by 100. We note that this tends to be unfavourable to our method, which has zero set-up time, unlike the direct method (which needs to compute a decomposition) or AMG (which needs to setup the preconditioner).

Recall that $1 \leq s(q) \leq n$, where $n$ is the number of nodes of the graph, and that $s(q)$ is the average number of roots alg. 1 outputs. Generally, the higher $s(q)$ is, the faster our algorithm. 
$s(q)$ will of course vary depending on the graph, and so in the comparisons we pick a range that is appropriate for each graph. We set the range such that $s(q)$ would vary approximately between 1\% and 50\% of $n$, the number of nodes. We picked 8 values on a logarithmic scale. 

The graphs we tested are as follows:
\begin{itemize}
\item ``circle'' : a ring graph of size $27,000$
\item ``grid\_2d'': a 2D lattice of size $164 \times 164=26,896$
\item ``grid\_3d'': a 3D lattice of size $30^3 = 27,000$
\item ``barabasi\_albert'': A Barabasi-Albert random graph with $n = 3000$ and $k = 30$ (average degree)
\item ``noisy\_heart``: a k-nearest neighbour graph obtained from $n = 4096$ points sampled from the parametric surface $x = \sin(\theta) \cos(\phi),\sin(\theta) y = \sin(\phi) (1+\exp(-0.1\theta)),z=\cos(\theta)(0.1+\theta)$ for $\theta \in [0,\pi], \phi \in [0,2\pi]$. We added a small random Gaussian offset to each point, and the surface looks heart-shaped when plotted, hence the name. 
\end{itemize}
Results are shown in Fig.~\ref{fig:results}. We plot run-time as a function of $s(q)$, to ease comparison across graphs. Our method is competitive compared to a direct solver for a range of values of $q$. Iterative methods make a relatively poor showing here, but they are expected to scale better with $n$. Also, we need to solve for several right-hand sides, and block CG methods may be more appropriate \cite{o1980block}. 
Finally, we have also checked that our algorithm scales to very large graphs. On a Barabasi-Albert random graph of size $n=1,000,000$ and 40 links per node, running our algorithm even at low $q=6\cdot10^{-3}$ (corresponding to $s(q) \approx 100$) takes a very reasonable 1/5 sec per realisation. 
\vspace{.3cm}

\noindent {\large \bf 4 \quad Discussion}
\vspace{.2cm}
%\section{Discussion}
%\label{sec:discussion}

Random forests on graphs lead to simple estimators for inverse traces of diagonally dominant matrices, and we find good practical performance. The small memory footprint is especially notable (all quantities stored scale in $\O(n)$). There are also several promising avenues for improvement. In many scenarios, what is needed is to evaluate $s(q)$ for a range of values of $q$, and the ``coupled forests'' algorithm of \cite{avena_two_2017} can be use to directly estimate $s(q)$ over a range much more cheaply than by running independent forests for a grid of values. The method can also be extended to estimate the values on the diagonal of $(\bL + q\bI)^{-1}$, a refinement we will describe in future work.

%\bibliographystyle{plain}

%\bibliography{ref}

\begin{thebibliography}{10}

\bibitem{anderson2010introduction}
Greg~W Anderson, Alice Guionnet, and Ofer Zeitouni.
\newblock {\em An introduction to random matrices}, volume 118.
\newblock Cambridge university press, 2010.

\bibitem{avena2013some}
L~Avena and A~Gaudilliere.
\newblock On some random forests with determinantal roots.
\newblock {\em arXiv preprint arXiv:1310.1723}, 2013.

\bibitem{avena_two_2017}
L.~Avena and A.~Gaudillière.
\newblock Two {Applications} of {Random} {Spanning} {Forests}.
\newblock {\em Journal of Theoretical Probability}, July 2017.

\bibitem{avena2018random}
Luca Avena, Fabienne Castell, Alexandre Gaudilli{\`e}re, and Clothilde
  M{\'e}lot.
\newblock Random forests and networks analysis.
\newblock {\em Journal of Statistical Physics}, 173(3-4):985--1027, 2018.

\bibitem{avron_randomized_2011-1}
Haim Avron and Sivan Toledo.
\newblock Randomized algorithms for estimating the trace of an implicit
  symmetric positive semi-definite matrix.
\newblock {\em Journal of the ACM}, 58(2):1--34, April 2011.

\bibitem{barrett1994templates}
Richard Barrett, Michael~W Berry, Tony~F Chan, James Demmel, June Donato, Jack
  Dongarra, Victor Eijkhout, Roldan Pozo, Charles Romine, and Henk Van~der
  Vorst.
\newblock {\em Templates for the solution of linear systems: building blocks
  for iterative methods}, volume~43.
\newblock Siam, 1994.

\bibitem{chung_complex_2006}
Fan~RK Chung and Linyuan Lu.
\newblock {\em Complex graphs and networks}, volume 107.
\newblock American mathematical society Providence, 2006.

\bibitem{girard1989fast}
A~Girard.
\newblock A fast ‘monte-carlo cross-validation’procedure for large least
  squares problems with noisy data.
\newblock {\em Numerische Mathematik}, 56(1):1--23, 1989.

\bibitem{girard1987algorithme}
Didier Girard.
\newblock Un algorithme simple et rapide pour la validation crois{\'e}e
  g{\'e}n{\'e}ralis{\'e}e sur des probl{\`e}mes de grande taille.
\newblock Technical report, 1987.

\bibitem{ESL}
Trevor Hastie, Robert Tibshirani, and Jerome Friedman.
\newblock {\em The elements of statistical learning}.
\newblock Springer, 2009.

\bibitem{hunter2014computing}
Timothy Hunter, Ahmed~El Alaoui, and Alexandre Bayen.
\newblock Computing the log-determinant of symmetric, diagonally dominant
  matrices in near-linear time.
\newblock {\em arXiv preprint arXiv:1408.1693}, 2014.

\bibitem{hutchinson1990stochastic}
Michael~F Hutchinson.
\newblock A stochastic estimator of the trace of the influence matrix for
  laplacian smoothing splines.
\newblock {\em Communications in Statistics-Simulation and Computation},
  19(2):433--450, 1990.

\bibitem{kelner2013simple}
Jonathan~A Kelner, Lorenzo Orecchia, Aaron Sidford, and Zeyuan~Allen Zhu.
\newblock A simple, combinatorial algorithm for solving sdd systems in
  nearly-linear time.
\newblock In {\em Proceedings of the forty-fifth annual ACM symposium on Theory
  of computing}, pages 911--920. ACM, 2013.

\bibitem{mahoney2011randomized}
Michael~W Mahoney et~al.
\newblock Randomized algorithms for matrices and data.
\newblock {\em Foundations and Trends{\textregistered} in Machine Learning},
  3(2):123--224, 2011.

\bibitem{o1980block}
Dianne~P O'Leary.
\newblock The block conjugate gradient algorithm and related methods.
\newblock {\em Linear algebra and its applications}, 29:293--322, 1980.

\bibitem{rue2005gaussian}
Havard Rue and Leonhard Held.
\newblock {\em Gaussian Markov random fields: theory and applications}.
\newblock Chapman and Hall/CRC, 2005.

\bibitem{ruge1987algebraic}
John~W Ruge and Klaus St{\"u}ben.
\newblock Algebraic multigrid.
\newblock In {\em Multigrid methods}, pages 73--130. SIAM, 1987.

\bibitem{stein2013stochastic}
Michael~L Stein, Jie Chen, Mihai Anitescu, et~al.
\newblock Stochastic approximation of score functions for gaussian processes.
\newblock {\em The Annals of Applied Statistics}, 7(2):1162--1191, 2013.

\bibitem{wilson1996generating}
David~Bruce Wilson.
\newblock Generating random spanning trees more quickly than the cover time.
\newblock In {\em Proceedings of the Twenty-eighth Annual ACM Symposium on the
  Theory of Computing (STOC)}, volume~96, pages 296--303. Citeseer, 1996.

\end{thebibliography}

\end{document}